\documentclass{iaus}
\usepackage{graphicx}

\title[IAUS 271.~~Simulations of Fully Convective Stars] 
{Differential Rotation and Magnetism in Simulations of Fully Convective Stars}

\author[Matthew K. Browning]   
{Matthew K. Browning$^1$}

\affiliation{$^1$Canadian Institute for Theoretical Astrophysics,
  \\ University of Toronto,
60 St. George St, Toronto, Canada \\ email: {\tt browning@cita.utoronto.ca} }

\pubyear{2010}
\volume{271}  
\pagerange{119--126}
\jname{Astrophysical Dynamics from Stars to Galaxies}
\editors{N. Brummell \& A.S. Brun, eds.}
\begin{document}

\maketitle

\begin{abstract}
  Stars of sufficiently low mass are convective throughout their interiors,
  and so do not possess an internal boundary layer akin to the solar
  tachocline. Because that interface figures so prominently in many
  theories of the solar magnetic dynamo, a widespread expectation had been
  that fully convective stars would exhibit surface magnetic behavior very
  different from that realized in more massive stars.  Here I describe how
  recent observations and theoretical models of dynamo action in low-mass
  stars are partly confirming, and partly confounding, this basic
  expectation.  In particular, I present the results of 3--D MHD
  simulations of dynamo action by convection in rotating spherical shells
  that approximate the interiors of 0.3 solar-mass stars at a range of
  rotation rates.  The simulated stars can establish latitudinal differential
  rotation at their surfaces which is solar-like at ``rapid'' rotation
  rates (defined within) and anti-solar at slower rotation rates; the
  differential rotation is greatly reduced by feedback from strong
  dynamo-generated magnetic fields in some parameter regimes.  I argue that
  this ``flip'' in the sense of differential rotation may be observable in
  the near future.  I also briefly describe how the strength and morphology
  of the magnetic fields varies with the rotation rate of the simulated
  star, and show that the maximum magnetic energies attained are compatible
  with simple scaling arguments.

\end{abstract}

\firstsection 
\section{Introduction: Puzzles of Low-Mass Stellar Magnetism}

Magnetic fields are ubiquitous in low-mass stars, and in at
least some cases those magnetic fields exhibit a remarkable amount of
spatial and temporal organization.  The most famous example is the Sun's
cyclical magnetism: sunspots appear on the solar disk first at
mid-latitudes, then progressively nearer the equator over the course of a
roughly 11-year cycle; the number and polarity of the spots varies in the
same way (see., e.g., \cite[Ossendrijver 2003]{Ossendrijver2003}).  These
organized magnetic fields are widely (though not universally) thought to be
generated partly in the tachocline of shear at the base of the solar
convective envelope -- in part because it is a site of strong differential
rotation, but also because the stable stratification below the convection
zone might allow magnetic fields to be amplified enormously before becoming
susceptible to magnetic buoyancy instabilities (see, e.g., \cite[Parker
  1993]{Parker1993}; see \cite[Spruit 2010]{Spruit2010} for a different
view).

But not all stars have a tachocline.  Moving down the main sequence to
lower masses and cooler temperatures, the convective envelope deepens and the
radiative core shrinks.  Stars of less than about a third a solar mass
(corresponding to spectral types of roughly M3.5 or later) are thought to
be convective throughout their interiors, and so do not possess an internal
boundary layer akin to the solar tachocline.  Although these low-mass stars
might in principle still possess differential rotation -- like the
latitudinal shear present throughout the solar convection zone -- a
widespread theoretical expectation had been that they would exhibit
magnetic fields that differed appreciably from those realized in more
massive solar-like stars (e.g., \cite[Durney et
  al. 1993]{Durney_etal1993}).

Observations of stellar magnetism paint a somewhat murkier picture.  On the
one hand, there have been recent suggestions (particularly from
spectropolarimetric observations) that the surface magnetic topologies of
stars with a small radiative core \emph{do} differ appreciably from those
of stars that are fully convective (e.g., \cite[Morin et
  al. 2010]{Morin_etal2010}; \cite[Reiners \& Basri
  2009]{Reiners_Basri2009}).  Fully convective stars (with late-M spectral
types) also appear to spin down (through angular momentum loss via a
magnetized stellar wind) much more slowly than early-M dwarfs (e.g.,
\cite[Browning et al. 2010]{Browning_etal2010}; \cite[Reiners et
  al. 2009]{Reiners_etal2009}), though it is unclear whether this is due to
changes in the strength or morphology of surface magnetic fields or to
changes in the mass loss rate.  On the other hand, it also seems clear that
some fully convective stars can build magnetic fields with remarkably
strong large-scale components (\cite[Donati et al. 2006]{Donati_etal2006}),
with the strength of those fields sensitive to rotation at some level
(e.g., \cite[Mohanty \& Basri 2003]{Mohanty_Basri2003}).  Indeed, some
recent comparisons between line-of-sight and unsigned field measurements
(extracted from Stokes V and Stokes I observations) suggest that some fully
convective stars actually harbor \emph{more} organized fields than slightly
more massive stars with a small radiative core (\cite[Reiners \& Basri
  2009]{Reiners_Basri2009}).  Puzzlingly, some of these stars appear to
show large-scale magnetic field organization but no evident surface
differential rotation (e.g., \cite[Donati et al. 2006]{Donati_etal2006}).

The most central question raised by these observations is, simply, how are
spatially organized fields realized in fully convective stars?  But this
basic issue is tightly linked to a whole set of other questions: what is
the nature of the differential rotation in these stars, and what role does
it play in the dynamo process?  Can these stars ever drive solar-like
differential rotation at observable levels?  How do the strength and
morphology of the dynamo-generated fields vary with rotation rate?  How
does the magnetism modify the transport of energy and angular momentum
throught the stellar interior?  These questions motivate the work described
in this paper.

Here, I describe the results of a series of 3-D simulations of convection
and magnetism in rotating spherical domains that are intended to represent
fully convective stars of 0.3 solar masses at various rotation rates.  In
\S 2 I describe the basic setup of these simulations and the numerical
methods employed.  The convective flow patterns and some aspects of energy
transport are detailed in \S 3.  I describe the differential rotation that
arises in these models in \S 4.  In \S 5 I describe the strength and
morphology of the magnetic fields realized at various rotation rates.

\section{Model Formulation and Numerical Methods}

The simulations described here are highly idealized representations of 0.3
solar-mass stars rotating between one-tenth and ten times as rapidly as the
Sun (0.1--10 $\times \Omega_{0} = 2.6 \times 10^{-6}$ s$^{-1}$).  They were
all carried out using the Anelastic Spherical Harmonic (ASH) code, which
solves the 3--D Navier-Stokes equations with magnetism in the anelastic
approximation (\cite[Clune et al. 1999]{Clune_etal1999}; \cite[Miesch et
  al. 2000]{Miesch_etal2000}; \cite[Brun, Miesch \& Toomre
  2004]{BMT_2004}).  The setup at each rotation rate is essentially
identical to that described in \cite[Browning (2008)]{Browning_2008}; here
I summarize only the most important details.  Before delving into them,
note that the ASH code is well-suited to this problem because it correctly
captures the global spherical geometry of the star, thereby allowing the
study of intrinsically large-scale processes like differential rotation,
meridional circulation, and global-scale dynamo action.  The tradeoff, of
course, is that because we can only resolve a finite range of spatial and
temporal scales in the simulation, including the largest possible length
scales (the radius of the star) in our modeling implies that the smallest
lengths resolved are still quite large -- typically about 1 Mm.  Thus I
focus here on the dynamics of the largest scales, while recognizing that at
some level of detail these must be influenced by smaller-scale dynamics
that these simulations cannot resolve.

The spherical computational domain typically extends from 0.06-0.96R, where
$R$ is the overall stellar radius of $2.07 \times 10^{10}$ cm, thus
excluding both the surface boundary layer and the innermost portions of the
star. I exclude the inner few percent of the star from these calculations
both because the coordinate systems employed in ASH are singular there, and
because the small numerical mesh sizes at the center of the star would
require impractically small timesteps.  The initial stratifications of the
mean density, energy generation rate, gravity, radiative diffusivity, and
entropy gradient $dS/dr$ are adopted from a 1-D stellar model (I. Baraffe,
private communication, after \cite[Baraffe \& Chabrier
  2003]{Baraffe_Chabrier2003}).  These thermodynamic quantities are updated
throughout the course of the simulation as the evolving convection modifies
the spherically symmetric mean state.  Variables are expanded in terms of
spherical harmonic basis functions $Y_l^m(\theta, \phi)$ in the horizontal
directions and Chebyshev polynomials $T_n(r)$ in the radial. As with all numerical simulations,
the eddy viscosities and diffusivities employed are vastly greater than
their counterparts in actual stars; here I have taken these to be constant
in radius, and adopted a Prandtl number $\nu/\kappa=0.25$ and a magnetic
Prandtl number $Pm=\nu/\eta$ that varies between $0.25$ and $8$ depending
on the simulation.  At a detailed level, the flows and magnetic fields
attained in the simulations are sensitive to the values of these
non-dimensional numbers.  Because of this, I focus here on the broad trends
these simulations exhibit: the types of flows and magnetic fields they can
achieve as the basic parameters of the problem are varied, rather than the
precise values of magnetic energy, zonal wind velocity, etc, that are
attained.  I particularly concentrate here on the role that changes in the
overall stellar rotation rate play, since this turns out to have an
outsized influence on the flows and magnetic fields that are achieved.

\section{Convective Flows and Energy Transport}

\begin{figure}[b]
\begin{center}
 \includegraphics[width=5.5in]{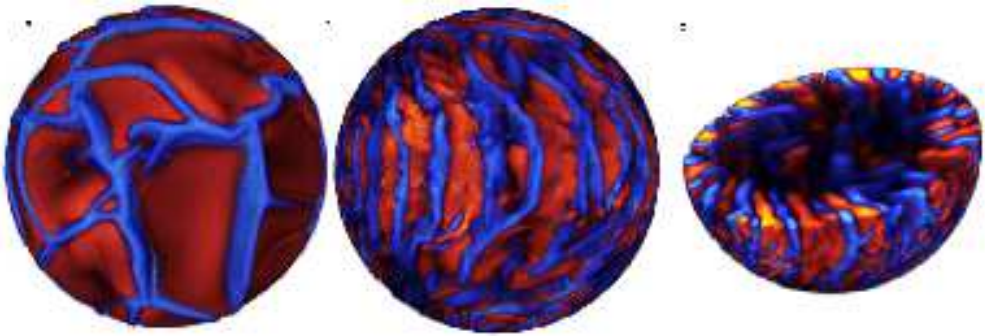} 
 \caption{Radial velocity in simulations rotating at the solar
   angular velocity ($b$, $c$) and tenfold slower ($a$).  Upflows are
   reddish (light), downflows are blueish (dark).  Panel $c$ shows a
   cutaway of one hemisphere. }
   \label{fig1}
\end{center}
\end{figure}

The convective flows in these simulations possess structure on many scales.
Although many small-scale features continually emerge, propagate, and
survive only for a short while, there are also large-scale organized
motions that can persist for extended intervals.  A sampling of this
behavior is provided by Figure 1, which shows volume renderings of the
radial velocity in two simulations, one rotating at the solar angular
velocity and the other ten times slower.  (The opacity mapping used in that
figure is such that only motions near the outer boundary of the simulation
are visible in Fig. 1$a$, 1$b$; Figure 1$c$ shows a cutaway of one
hemisphere in the more rapidly rotating simulation, to highlight the radial
extent of the motions.)  Rotation has a strong effect on the convective
patterns that are realized: when rotation is very slow, the convection is
more or less isotropic on each spherical surface, with a network of upflows
and downflows that meanders in orientation over the sphere.  Increasing the
rotation rate breaks this symmetry and imposes a preferred direction,
leading (in some cases) to convective rolls like those shown in Figure
1$b$, 1$c$.  (If the eddy viscosity in the simulations is decreased, these
rolls become less pronounced and can break up, but there is still a
tendency towards alignment along the rotation axis.)  These roll-like
structures are a well-known feature of convection in rotating spherical
shells (see, e.g., \cite[Busse 2002]{Busse_2002}; \cite[Gilman
  1977]{Gilman_1977}).

One other feature worth highlighting is the strong asymmetry between
upflows and downflows realized near the top of the simulated stars: the
downflows are stronger and narrower than the former, mainly because of the
strong density stratification.  (Downflows are cool and contract; upflows
are hot and expand.)  Deeper in the interior, the flows are weaker and of
somewhat larger physical scale: motions can span large fractions of a
hemisphere and extend radially for some distance.  Flow amplitudes also
vary appreciably with depth, with typical rms velocities declining by a
factor of about ten in going from the surface to the center.

The variation in flow amplitude with radius is linked to both the density
stratification and to radial variations in the amount of energy that must
be transported by convection.  Convection ultimately arises because of a
need to transport heat outwards: if more energy has to be carried by
convection, the convective velocity will generally be higher.  Although the
interior is unstably stratified everywhere, radiation still carries some of
the energy at small radii.  This is because the end state of efficient
convection is an interior stratification that is approximately isentropic,
\emph{not} isothermal as in unstratified convection: thus there is still a
non-zero radiative flux.  Together with the overall increase of the total
luminosity with radius (out to the point where nuclear energy generation
stops), this implies that the total luminosity carried by convection peaks
at large radii (around $r=0.80R$). Thus the convective velocity is
appreciably greater near the surface than at depth.  Another important
effect arises because of the asymmetry between upflows and downflows: this
implies a negative (inward) kinetic energy flux, which in a steady state
must be compensated for by an increased outward enthalpy (convective)
flux. This effect, too, depends on depth (since the star is more strongly
stratified near the surface), again leading to more vigorous convection
near the top of the simulation domain.

\section{Differential Rotation: Solar or Anti-Solar}

In addition to transporting heat, the convective flows also transport
angular momentum.  One might naively expect that parcels of fluid would
\emph{individually} conserve angular momentum -- so that (for instance) a fluid element
moving outward would tend to slow down (move retrograde relative to the
frame), and a fluid element moving latitudinally from equator to pole would
tend to spin up.  This would imply anti-solar differential rotation at the
surface, with a slow equator and fast poles, and longitudinal velocities
that decrease with distance from the rotation axis.  The fact that the Sun
in fact has a fast equator and slow pole is enough to suggest that there is
more to the story than this: that convection, acting in concert or conflict
with meridional circulations and magnetic fields, can redistribute angular
momentum in surprising ways.  Here I describe briefly the types of
differential rotation that are achieved in these simulations at varying
rotation rates, while deferring an in-depth analysis of the angular
momentum transport to forthcoming work (\cite[Browning \& Miesch
  2010]{Browning_Miesch2010}). 

All of the MHD calculations described here had hydrodynamical progenitor
simulations.  These all began in a state of uniform rotation, but
convection quickly established interior rotation profiles that varied with
radius and latitude.  The resulting differential rotation, displayed in
Figure 2 for two \emph{hydrodynamic} cases, depends on the overall stellar
rotation rate: in Figure 2$a$, from a simulation rotating at one-tenth the
solar angular velocity, the rotation profile is ``anti-solar'' at the
surface, with a slow equator and fast poles; in Figure 2$b$, from a
simulation rotating at the solar angular velocity, the rotation profile is
solar-like at the surface.  Both cases also show angular velocity contrasts
in radius, with angular velocity decreasing with depth in the more rapidly
rotating case and decreasing with depth in the slower rotator.  In the
simulation rotating at the solar rate, the interior angular velocity
profile is largely constant on cylinders, reflecting the strong
Proudman-Taylor constraint; in the more slowly rotating case this
cylindrical alignment is not evident.

Similar transitions in the nature of angular momentum transport and
differential rotation have been noted by several previous authors in other
contexts.  In particular, Gilman (1977) noted that his simulations of a
solar-like convection zone exhibited solar-like equatorial acceleration in
some regimes and anti-solar rotation profiles in others; he identified the
transition between these two regimes with a transition from a convective
Rossby number (essentially the ratio of buoyancy driving to Coriolis
forces) greater than unity to less than unity.  When the
rotational influence on the convection was strong relative to buoyancy
driving, he attained equatorial acceleration; when it was weak, the equator
rotated more slowly than the poles.  Similar results were obtained in
simulations of core convection in A-type stars (\cite[Browning et
  al. 2004]{Browning_etal2004}). The simulations here exhibit qualitatively
similar behavior.  Convection influenced by rotation (that is, with a
Rossby number roughly less than unity) tends to drive solar-like
differential rotation; when rotation is slower, the differential rotation
profile is anti-solar.

\begin{figure}[b]
\begin{center}
 \includegraphics[width=3.2in]{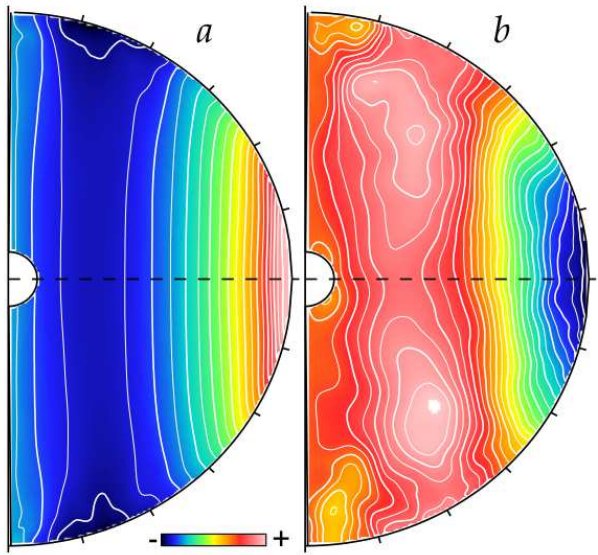} 
 \caption{Differential rotation achieved in sample hydrodynamic simulations
   rotating at the solar angular velocity ($a$) and ten times slower
   ($b$).  Light tones are prograde, dark ones retrograde.}
   \label{fig2}
\end{center}
\end{figure}

The interior rotation profiles can be quite different in the presence of
strong dynamo-generated magnetic fields.  In MHD simulations, the magnetic
fields react back strongly upon the flows, acting to quench and in some
instances essentially eliminate the differential rotation.  An example of
this for a simulation rotating at the solar rate was explored in
\cite[Browning (2008)]{Browning_2008}.  The extent to which the
differential rotation is reduced depends on the overall magnetic field
strength, which in turn (as discussed below) depends somewhat on the
overall stellar rotation rate.  Thus, there is a ``sweet spot'' for
differential rotation: when the simulations rotate very slowly, they drive
weak anti-solar differential rotation; when the rotation rate is somewhat larger they drive
strong solar-like differential rotation; and when rotation is more rapid
still they build magnetic fields so strong that the differential rotation
is partly quenched.  These transitions occur at fairly slow overall rotation
rates: because the luminosities of M-dwarfs are so low, their convective
velocities are also low, and so they are strongly influenced by rotation
even at solar-like angular velocities.  Only the very slowest rotators
(those with rotation periods much longer than the solar value) drive
anti-solar differential rotation.

It seems plausible that these basic predictions could be testable in the
near future by photometric monitoring with, e.g., Kepler.  The exact
rotation periods at which stars might transition from anti-solar to
solar-like differential rotation, and then again to no evident differential
rotation (because it has been wiped out by the dynamo-generated Maxwell
stresses), are probably not reliably predictable by these calculations.
They depend on the values of viscosity, magnetic diffusivity, etc, adopted
in the modeling. But the basic existence of these different regimes, and
their rough dependence on rotation rate, might well be robust.  The main
challenge observationally will probably be to measure differential rotation
reliably in the slowest rotators, which are presumably less active and so less
spotted. 

\section{Dynamo Action Achieved}

The flows in each simulation act as a magnetic dynamo, amplifying a tiny
seed field by many orders of magnitude and sustaining it against Ohmic
decay. The magnetic energy (ME) grows exponentially until it reaches a
steady state.  In the most rapid rotators (which, again, are not really
rotating all that rapidly in absolute terms), the final magnetic energy
density is approximately in equipartition with the flows; in slower
rotators it tends to be smaller.  The exact values of the magnetic energy
are sensitive at some level to the values of the magnetic Reynolds number,
magnetic Prandtl number, etc, that I have adopted in each calculation.  To
give a specific example: in an evolved calculation rotating at the solar
angular velocity, ME was approximately 120\% of the total kinetic energy KE
(relative to the rotating frame) and about 140\% of the convective
(non-axisymmetric) kinetic energy (CKE).

As the fields grow, they react back on the flows through the Lorentz force.
Thus KE begins to decline once ME reaches a threshold value of about 5\% of
KE; here this decline is associated mainly with a decrease in the energy of
differential rotation DRKE, whereas CKE is less affected by the
growing fields.

Like the flows that build them, the magnetic fields possess both intricate
small-scale structure and substantial large-scale components.  The typical
length scale of the field increases with depth, partly tracking the radial
variation in the size of typical convective flows.  The smallest field
structures are typically on finer scales than the smallest flow fields,
partly because I have adopted a magnetic Prandtl number Pm greater than
unity.  By decomposing the magnetism into its azimuthal mean (TME), and
fluctuations around that mean (FME), we can gain a coarse estimate of the
typical size of field structures: if the field is predominantly on small
scales, only a small signal will survive this azimuthal averaging.  In
these simulations, TME accounts for at most about 20\% of the total magnetic energy
in the bulk of the interior; it is smallest near the surface (where TME is
typically less than 5 \% ME), and largest (as a fraction of ME) at depth.

The fraction of energy in the mean (axisymmetric) components increases with
increasing rotation rate.  The ratio of the toroidal mean energy to the
poloidal mean also changes: in the very slowest rotators the two components
are comparable, while in more rapidly rotating cases with strong
differential rotation, TME exceeds the poloidal mean energy (PME) by a
factor of a few.

The mean (axisymmetric) fields realized in some of the simulations are
remarkably strong and long-lived.  Mean toroidal field strengths can exceed
10 kG in some locations; some prominent field structures persist for
thousands of days.  The overall field polarity is stable over long
intervals (decades), in sharp contrast to some simulations of solar
convection without a tachocline (\cite[Brun et al. 2004]{BMT_2004}), in
which the field polarity flipped at irregular intervals of less than 600
days.

\section{Closing Thoughts}

The overall picture that emerges from these simulations is that fully
convective stars can act effectively as magnetic dynamos, building fields
that have structure on both large and small spatial scales.  The
large-scale fields can be remarkably strong and long-lived.  The convection
drives differential rotation, in a manner that depends on the overall
rotation rate: the very slowest rotators would appear anti-solar at their
surface, while more rapid rotators establish a fast equator and slow pole.
That differential rotation is, however, largely quenched by Maxwell stresses in
cases that build strong magnetic fields.

The maximum magnetic energy densities attained in these calculations are of
order equipartition with the kinetic energy density relative to the
rotating frame.  Although stronger magnetic fields might be possible in
some instances (see, e.g., \cite[Featherstone et
  al. 2009]{Featherstone_etal2009}), it is worth noting that the assumption
of equipartition yields magnetic field estimates broadly in line with those
recently predicted on somewhat different grounds by Christensen and collaborators
(see, e.g., \cite[Christensen et al. 2009]{Christensen_etal2009};
\cite[Reiners et al. 2009]{Reiners_etal2009}).  Specifically, Reiners et
al. (2009) argue, based on an energy flux scaling derived empirically from
a series of planetary dynamo calculations, that the magnetic flux in stars
should scale approximately as
\begin{equation}
  Bf \propto  M^{1/6} L^{1/3} R^{-7/6}
\end{equation}
with $M$, $L$, and $R$ the stellar mass, luminosity and radius, and where
$Bf$ is the surface magnetic flux.  This turns out to be the same
scaling one derives by assuming that $1.$ magnetic fields are in
approximate equipartition with the convective kinetic energy and $2.$ that
convective energy is given roughly by mixing-length arguments, so that it
is proportional to the heat flux to the one-third power.  To order of
magnitude, the convective luminosity is given by
\begin{equation}
  L \sim \frac{\textrm{convective energy}}{\textrm{convective overturning
      time}} \sim \frac{ \rho v^2 (\frac{4}{3} \pi d^3)}{l_{c} / v}
\end{equation}
with $d$ a lengthscale characterizing the depth of the convection zone and
$l_c$ the length characterizing convective eddies.  This in turn implies that
$ v \sim (\frac{3L}{4 \pi \rho} l_c)^{1/3} \frac{1}{d} $.
Meanwhile, equipartition of kinetic and magnetic energy densities implies
that $B \sim \rho^{1/2} v$.  If we then take $\rho \sim M/R^{3}$, and
assume that to order of magnitude all lengthscales are comparable ($l_c
\sim d \sim R$), we obtain that $v \sim L^{1/3} R^{1/3} M^{-1/3}$, so that
finally $B \sim L^{1/3} M^{1/6} R^{-7/6}$, as in the Reiners et al. (2009)
scaling.  This is not to suggest that greater or lesser field energies are
not possible, but does indicate that objects deviating greatly from the
Reiners et al. (2009) scaling are breaking one of the assumptions I made
above.  Either equipartition does not hold, or the relevant flow velocity
is not related to the background heat flux in the way assumed here.

It is a pleasure to acknowledge many helpful conversations about this and
related puzzles in stellar rotation and magnetism with Juri and the other
members of the ``ASH mob.''

\begin{discussion}

 \discuss{A. Brandenburg}{ 
Is the dependence of rotation rate on spectral type real and how does
this depend on age?
}

\discuss{M. Browning}{ The conclusion that very low-mass stars take longer
  to spin down than somewhat higher-mass ones appears to me to be quite
  robust.  But there is certainly an age effect: if you look at young
  enough clusters, many stars of all spectral types are still rotating
  quite rapidly.  In the field, though -- i.e., looking at much older stars
  -- you tend to find that very few early-M stars are detectably rotating
  (see, e.g., \cite[Browning et al. (2010)]{Browning_etal2010}, where we
  used Keck HIRES spectra to look for signs of rotation in dozens of field
  M-dwarfs).  A significantly larger fraction of late-M
  stars (or L-dwarfs) are detectably rotating. }
  
   \discuss{K. Moffat}{ 
Your conclusion that rotation plus convection are conducive to the
growth of large-scale fields is hardly new! These ingredients imply
large-scale helicity and so a corresponding $\alpha$-effect. Have you
been able to interpret your simulations from this ``mean-field'' point
of view?
}

\discuss{M. Browning}{ The short answer is no, these simulations do not
  seem to be particularly well-described by mean-field theory.  There is
  certainly large-scale helicity of the predicted sense, but it doesn't
  seem to be especially well-linked to the growth of the large-scale field.
  The role of rotation in these simulations seems to be partly about
  increasing the correlation length of the convection to something on the
  order of the domain size, rather than just imparting a preferred sign to
  the helicity in each hemisphere.  That said, we certainly have more work
  to do in trying to make connections between these results and mean-field theory.}

 \discuss{K. Ferriere}{ 
Why in the Sun does the equator rotate faster than the poles?
}

\discuss{M. Browning}{ I would say that, honestly, we still have a ``description'' of
  how this happens rather than a first-principles theory.  (See Miesch et al.,
  these proceedings, for some of that description.)  Ultimately there are
  Reynolds stresses arising from correlations in the fluctuating velocity
  components, and the sense of those stresses is such that you break the
  tendency for individual parcels of fluid to just conserve angular
  momentum as they move outward or inward. But baroclinic effects,
  meridional circulations, etc, all appear to play roles as well.}


\end{discussion}


\begin{thebibliography}{}

\bibitem[Browning (2008)]{Browning_2008}{Browning, M.K., 2008,
  \textit{ApJ}, 676, 1262}
  
\bibitem[Browning et al. (2010)]{Browning_etal2010}{Browning, M.K., Basri,
  G., Marcy, G.W., West, A.A., \& Zhang, J., 2010, \textit{AJ}, 139, 504}

 \bibitem[Brun, Miesch \& Toomre (2004)]{BMT_2004}{ Brun, A.\ S., Miesch, M.\ S., \& Toomre, J. 2004, \textit{ApJ}, 614, 
   1073}

   \bibitem[Christensen et al. (2009)]{Christensen_etal2009}{Christensen,
     U.R., Holzwarth, V., \& Reiners, A., 2009, \textit{Nature}, 457, 167}

\bibitem[Donati et al. (2006)]{Donati_etal2006}{
  Donati, J.F., Forveille, T., Cameron, A.C., Barnes, J.R., Delfosse, X.,
  Jardine, M.M., \& Valenti, J.A., 2006, Science, 311, 633}
  
\bibitem[Durney et al. (1993)]{Durney_etal1993}{
  Durney, B.R., De Young, B.S., \& Roxburgh, I.W., 1993,
 \textit{Sol. Phys.}, 145, 207}

  \bibitem[Featherstone et al. (2009)]{Featherstone_etal2009}{Featherstone,
    N.A., Browning, M.K., Brun, A.S., \& Toomre, J., 2009, \textit{ApJ},
    706, 1000}

\bibitem[Gilman (1977)]{Gilman_1977}{Gilman, P.A., 1977, \textit{GAPFD}, 8, 93}  

\bibitem[Miesch et al. (2000)]{Miesch_etal2000}{ Miesch, M.\ S., Elliott,
  J.\ R., Toomre, J., Clune, T.\ L., Gl atzmaier, G.\ A., \& Gilman,
  P.\ A., 2000, \textit{ApJ}, 532, 593}


\bibitem[Mohanty \& Basri (2003)]{Mohanty_Basri2003}{
  Mohanty, S. \& Basri, G., 2003, \textit{ApJ}, 583, 451}

  \bibitem[Morin et al. (2010)]{Morin_etal2010}{ Morin, J., et al., 2010,
    \textit{MNRAS} 1077}

\bibitem[Ossendrrijver (2003)]{Osssendrijver_2003}
  {Ossendrijver, M. 2003}, \textit{Astron.\ Astrophys.\ Rev.}, 11, 287


\bibitem[Parker (1993)]{Parker_1993}{ Parker, E.\ N. 1993, \textit{ApJ},
  408, 707      }

  \bibitem[Reiners \& Basri (2009)]{Reiners_Basri2009}{Reiners, A. \&
    Basri, G., 2009, \textit{A\&A}, 496, 787}

\bibitem[Reiners et al. (2009)]{Reiners_etal2009}{Reiners, A., Basri, G.,
  \& Christensen, U.R., 2009, \textit{ApJ}, 697, 373}    

\bibitem[Spruit (2010)]{Spruit_2010}{Spruit, H.C. 2010, arXiv:1004.4545}

  
  
\end{thebibliography}
\end{document}